\documentclass[journal]{IEEEtran} % double column version
\usepackage{color}
\usepackage{cite}
\usepackage{amssymb,amsfonts,url}
\usepackage{algorithmic}
\usepackage{textcomp}
\usepackage[rgb]{xcolor}
\usepackage{float}
\usepackage{placeins}
\usepackage{psfrag}
\newcommand{\stz}{\rule{0mm}{2.3ex}}
\newcommand{\stzh}{\rule{0mm}{4ex}}
\usepackage{amsmath,graphicx,bm}
\usepackage{color,soul,cite}
\usepackage{MnSymbol}
\usepackage{multirow}
\usepackage{makecell}
\usepackage{xcolor,colortbl,hhline}
\usepackage[super]{nth}
\usepackage{subcaption}
\usepackage[colorlinks=true,linkcolor=black,anchorcolor=black,citecolor=black,urlcolor=blue]{hyperref}
\usepackage{setspace}
\usepackage{dblfloatfix,booktabs,caption}
\usepackage{algorithm}
\usepackage{algorithmic}
\newcommand{\texg}{\textcolor{green}}
\newcommand{\texr}{\textcolor{red}}
\ifCLASSINFOpdf
\else
\fi
\begin{document}
\title{Coded Speech Quality Measurement\\ by a Non-Intrusive PESQ-DNN}

\author{Ziyi Xu,
        Ziyue Zhao
        and Tim Fingscheidt,~\IEEEmembership{Senior Member,~IEEE}% <-this % stops a space
\thanks{Ziyi Xu and Tim Fingscheidt are with the Institute for Communications Technology, Technische Universit{\"a}t Braunschweig, 38106 Braunschweig, Germany (e-mails:$\left \{ \text{ziyi.xu, t.fingscheidt} \right \}$@tu-bs.de). Ziyue Zhao is with GN Hearing, 5612AB Eindhoven, The Netherlands (e-mail: zzhao@gnhearing.com).}}

\maketitle

\begin{abstract}
Wideband codecs such as AMR-WB or EVS are widely used in (mobile) speech communication. Evaluation of coded speech quality is often performed subjectively by an absolute category rating (ACR) listening test. However, the ACR test is impractical for online monitoring of speech communication networks. Perceptual evaluation of speech quality (PESQ) is one of the widely used metrics instrumentally predicting the results of an ACR test. However, the PESQ algorithm requires an original reference signal, which is usually unavailable in network monitoring, thus limiting its applicability. {\tt NISQA} is a new non-intrusive neural-network-based speech quality measure, focusing on super-wideband speech signals. In this work, however, we aim at predicting the well-known PESQ metric using a non-intrusive {\tt PESQ-DNN} model. We illustrate the potential of this model by predicting the PESQ scores of wideband-coded speech obtained from AMR-WB or EVS codecs operating at different bitrates in noisy, tandeming, and error-prone transmission conditions. We compare our methods with the state-of-the-art network topologies of {\tt QualityNet}, {\tt WaweNet}, and {\tt DNSMOS}---all applied to PESQ prediction---by measuring the mean absolute error (MAE) and the linear correlation coefficient (LCC). The proposed {\tt PESQ-DNN} offers the best total  MAE and LCC of 0.11 and 0.92, respectively, in conditions without frame loss, and still is best when including frame loss. Note that our model could be similarly used to non-intrusively predict POLQA or other (intrusive) metrics. Upon article acceptance, code will be provided at GitHub. 
\end{abstract}


\IEEEpeerreviewmaketitle

\section{Introduction}
%%%%%%%%%%%%%%%%%%%%%%%%%%%%%%%%%%%%%%%
Speech signals being processed by speech encoder, transmission system, and speech decoder are called coded speech. The speech encoder and decoder forming a speech codec aim to represent the speech signal in digital form with the smallest bitrate. However, the coded speech quality is typically impaired by far-end additive background noise, quantization noise, and transmission errors. Modern transmission systems, e.g., videoconferencing systems, digital cellular communication networks, and voice over internet protocol (VoIP), support wideband (WB) speech sampled at 16 kHz or even higher sampling rates. 

ITU-T G.722 \cite{ITUG722} was the first standardized wideband speech codec, operating at 64 kbps (or lower) and widely used in videoconferencing systems and cordless telephony. The adaptive multi-rate wideband (AMR-WB) codec \cite{AMRWB} is one of the most widely used mobile telephony speech codecs, offering nine operation modes with bitrates from 6.6 to 23.85 kbps.\linebreak
Since the introduction of AMR-WB to digital cellular services such as third-generation (3G) communication systems and beyond, superior speech quality and voice naturalness have pushed the expectation towards speech communication systems to a higher level \cite{bessette2002adaptive}. Compared to G.722 and AMR-WB, which only focus on speech signals, enhanced voice services (EVS) \cite{EVS} is a modern audio codec that enables high-definition (HD) quality for speech and music signals. EVS supports audio signals from narrowband (NB) to full-band (FB) and is implemented in the voice over LTE (VoLTE) services of the fourth-generation (4G) communication system and beyond. EVS offers 11 operation modes for WB speech signals with bitrates ranging from 5.9 to 96 kbps.

In the {\it design phase} of communication systems, evaluation of the quality of coded speech is usually performed by a subjective absolute category rating (ACR) listening test, in which naive human listeners (subjects) give their opinion about the perceived quality of each speech utterance with a score range from 1 (bad) to 5 (excellent). The averaged ACR scores over all subjects for an utterance or test condition is known as mean opinion score (MOS). This type of ACR test is typically performed in a laboratory environment to control the external influences on subjects' judgments following ITU-T P.800 \cite{ITUT_p800}. However, this controlled ACR test is time-consuming to prepare and conduct, and it is also costly to recruit naive subjects. 

For {\it online monitoring} of speech communication networks, e.g., to dynamically adjust the bitrate modes of the codecs based on the coded speech quality during operation, a subjective ACR test is impractical. Perceptual evaluation of speech quality (PESQ) \cite{ITUT_pesq_wb_corri}, perceptual objective listening quality assessment (POLQA) \cite{ITUT_polqa_2018}, and virtual speech quality objective listener (ViSQOL)\cite{hines2015visqol} are well-known instrumental metrics for evaluating (coded) speech quality. PESQ is designed to predict ACR listening test results and is widely used. PESQ, POLQA, and ViSQOL algorithms, however, require an original reference signal ({\it intrusive approaches}), which is used to estimate the perceived coded speech quality by measuring the perceptually weighted distance to the corresponding coded speech signal. However, such a reference signal is unavailable for network monitoring, thus limiting the applicability of PESQ, POLQA, and ViSQOL.

In recent years, data-driven approaches have attracted much attention in approximating highly non-linear functions even for estimating human perception. Abel et al.\ trained a support-vector-machine-based MOS predictor to intrusively predict subjective MOS scores for narrowband-to-wideband artificial speech bandwidth extension \cite{abel2016instrumental}. The development of deep neural networks (DNNs) pushes the performance even further and enables end-to-end training of speech quality DNNs to predict instrumental metrics or subjective listening test results even in a {\it non-intrusive} way, without the need for a reference signal \cite{soni2016novel,fu2018Inter,andersen2018nonintrusive,mittag2019non,avila2019non,gamper2019intrusive,catellier2020wawenets,dong2020attention,mittag2021nisqa,reddy2021dnsmos,xu2021inter,xu2021deepT,reddy2022dnsmos,xu2022iwaenc}. Soni et al.\ \cite{soni2016novel} use a simple autoencoder to extract features from the speech signal, which are used to train a single-layer neural network to predict the subjective quality scores for narrowband (NB) speech signals of fixed length. Fu et al.\ \cite{fu2018Inter} proposed a so-called {\tt QualityNet}, based on a bidirectional long short-term memory (BLSTM) structure, to predict PESQ scores from the amplitude spectrogram of the WB speech signal. Due to the BLSTM recurrent structure, {\tt QualityNet} can predict PESQ scores for speech signals with variable lengths. Focusing on predicting the subjective quality scores for transmitted super-wideband (SWB) speech signals, Mittag and M\"{o}ller proposed a BLSTM-based DNN dubbed as {\tt NISQA} in \cite{mittag2019non,mittag2021nisqa}. An advanced version of {\tt NISQA}, which predicts three orthogonal qualities (noisiness, coloration, and discontinuity) for fullband (FB) speech signals, is proposed in \cite{mittag2021nisqa}. Most non-intrusive speech quality prediction DNNs use the time-frequency (TF) domain representation of the speech signal as an input. However, Catellier et al.\ \cite{catellier2020wawenets} propose to directly predict PESQ scores from the waveform of the wideband speech signal with a convolutional neural network (CNN) called {\tt WaweNet}. Due to the lack of recurrent structures, {\tt WaweNet} can only deal with speech signals with a fixed length of three seconds. A long speech signal is cut first into several 3-second segments, and then the final estimation for this long speech signal is the averaged estimated PESQ scores across all the segments. 

In the Microsoft Deep Noise Suppression (DNS) Challenges \cite{reddy2021icassp,reddy2021interspeechIN}, Reddy et al. provided a DNN-based speech quality measure called {\tt DNSMOS} \cite{reddy2021dnsmos}. The {\tt DNSMOS} DNN model is specifically trained to predict subjective rating scores following ITU-T P.808 \cite{ITU_P808} for enhanced speech signals from DNS tasks. To offer a better overview of the perceived quality of the enhanced speech signal, the authors proposed an advanced version of {\tt DNSMOS} \cite{reddy2022dnsmos}, which separately estimates the qualities of speech component, background noise, and the overall enhanced speech following ITU-T P.835 \cite{ITUT_P835}. Both versions of {\tt DNSMOS} require the input speech signals having a fixed length of nine seconds. In our recent works \cite{xu2021inter,xu2021deepT,xu2022iwaenc}, we proposed an end-to-end {\tt PESQNet} {\it for DNS applications}, adapted from a BLSTM-based speech emotion recognition DNN \cite{Meyer2021}, to predict PESQ scores of the enhanced speech signal. In these works, the trained {\tt PESQNet} is employed as a mediator to provide a differentiable PESQ loss during a speech enhancement DNN training, aiming at maximizing the PESQ score of the enhanced speech signal. This method, however, works only if the DNS DNN and the {\tt PESQNet} speech quality predictor follow an alternating learning protocol, which is prohibitive in our current work: Here, we require a readily trained DNN for non-intrusive quality prediction of speech signals that have been transcoded by various different speech codecs and modes. 

Most speech quality prediction DNNs focus on estimating the perceptual quality for speech signals degraded by additive noise or enhanced signals obtained from different DNS methods. Only a few works, e.g., {\tt WaweNet} \cite{catellier2020wawenets} and both versions of {\tt NISQA} \cite{mittag2019non,mittag2021nisqa}, estimate the perceived quality for {\it coded} speech signals, while the latter one explicitly focuses on coded super-wideband (SWB) speech signals. 

In this work, we propose an end-to-end non-intrusive DNN model, which we call {\tt PESQ-DNN}, to explicitly estimate the PESQ scores for coded speech signals obtained from various WB codecs. We also consider the influence from different transmission conditions, including packet losses in error-prone transmission, far-end additive noise, and tandeming transmission, where two WB codecs are serially concatenated. To the best of our knowledge, such a non-intrusive PESQ estimation DNN explicitly developed for coded speech considering a wide range of realistic conditions has not yet been proposed. Furthermore, unlike {\tt WaweNet} or the {\tt DNSMOS} network, our proposed {\tt PESQ-DNN} is capable of estimating PESQ scores for coded speech signals of varying lengths. As most other scientific works, we do not follow the assessment procedure for machine learning models on speech quality estimation, standardized in ITU-T P.565 \cite{ITU_P565} and P.565.1 \cite{ITU_P565_1}. One reason for this is that these ITU-T recommendations do not regard the residual background noise condition, but we see it as very important. Furthermore, the models defined in the recommendations expect side information as input to the speech quality prediction model, such as a packet loss flag, which is, however, contradictory to our goal of predicting speech quality {\it just} on the basis of the measurement speech signal. We note that particularly in error-prone transmission conditions, we obviously aim at an even more challenging goal.

As concerns topology, we build upon \cite{xu2021deepT}, but many changes are required for the DNN to serve the speech communication monitoring needs targeted in this work: (1) Compared to {\tt PESQNet}, the novel {\tt PESQ-DNN} employs a complex spectrogram as input to explicitly consider phase influences in the perceived speech quality. Except for a few works, e.g., {\tt WaweNet}, most speech quality prediction DNNs employ amplitude or power spectrogram input, leading to the problem that speech quality degradations caused by phase distortions cannot be measured. (2) Inspired by \cite{mittag2021nisqa}, we introduce a self-attention pooling layer instead of the statistics pooling (average, standard deviation, minimum, and maximum) used in {\tt PESQNet}. The attention layer is inserted following the BLSTM layer, allowing the {\tt PESQ-DNN} to change focus dynamically on features obtained from different time instances, which is helpful to stabilize the training process and improve the inference performance. (3) Different to {\tt PESQNet}, we investigate different training losses, adapted from \cite{fu2018Inter}, to give physical meanings to the intermediate embeddings inside the {\tt PESQ-DNN}. Subsequently, the values of the intermediate embeddings are controlled in the investigated loss formulation and are supposed to stabilize the training progress. Except for introducing these modifications, a comprehensive ablation study illustrates the influence and value of the introduced modifications, thereby being another core contribution of this work.

The rest of the article is structured as follows: In Section\,\uppercase\expandafter{\romannumeral 2}, we introduce the signal model and our mathematical notations. We describe our proposed non-intrusive {\tt PESQ-DNN} in Section\,\uppercase\expandafter{\romannumeral 3}. Then we explain the experimental setup including the database, training, validation, and test conditions, baselines, and performance metrics in Section\,\uppercase\expandafter{\romannumeral 4}. Comprehensive ablation studies and discussions on results are given in Section\,\uppercase\expandafter{\romannumeral 5} and our work is concluded in Section\,\uppercase\expandafter{\romannumeral 6}.
%%%%%%%%%%%%%%%%%%%%%%%%%%%%%%%%%%%%%%%%
\section{Signal Model and Notations}
We assume the transmitter-sided speech encoder input signal $y(n)$ to comprise the original speech signal $s(n)$ and potentially residual\footnote{As $y(n)$ does {\it not} denote the microphone signal, but the speech encoder input signal, $d(n)$ models the {\it residual} noise after a potential noise suppression algorithm along with some slight speech distortions.} noise $d(n)$ as:
\begin{equation} \label{mic}
y(n)=s(n) + d(n),
\end{equation}
with $n$ being the discrete-time sample index. Afterwards, the signal $y(n)$ is processed by some speech encoder to obtain the bitstream, which is then passed through a transmission system and processed by the corresponding speech decoder. Finally, the coded speech obtained on the receiver side is denoted as $\hat{y}(n)$.

The proposed non-intrusive {\tt PESQ-DNN} estimates the PESQ scores of the coded speech signal $\hat{y}(n)$ based on its discrete Fourier transform (DFT) spectrogram $\hat{Y}_\ell(k)$, with frame index $\ell$, frequency bin index $k\!\in\!\mathcal{K}\!=\!\left \{0,1,\ldots,K\!-\!1\right \}$, and $K$ being the DFT size. This transformation is always performed via the fast Fourier transform (FFT), and successive FFT frames overlap in time.
%%%%%%%%%%%%%%%%%%%%%%%%%%%%%%%%%%%%%%%%%
\section{Proposed Non-Intrusive {\tt PESQ-DNN}}
\begin{figure}[t!]
	\centering
	\includegraphics[width=0.49\textwidth]{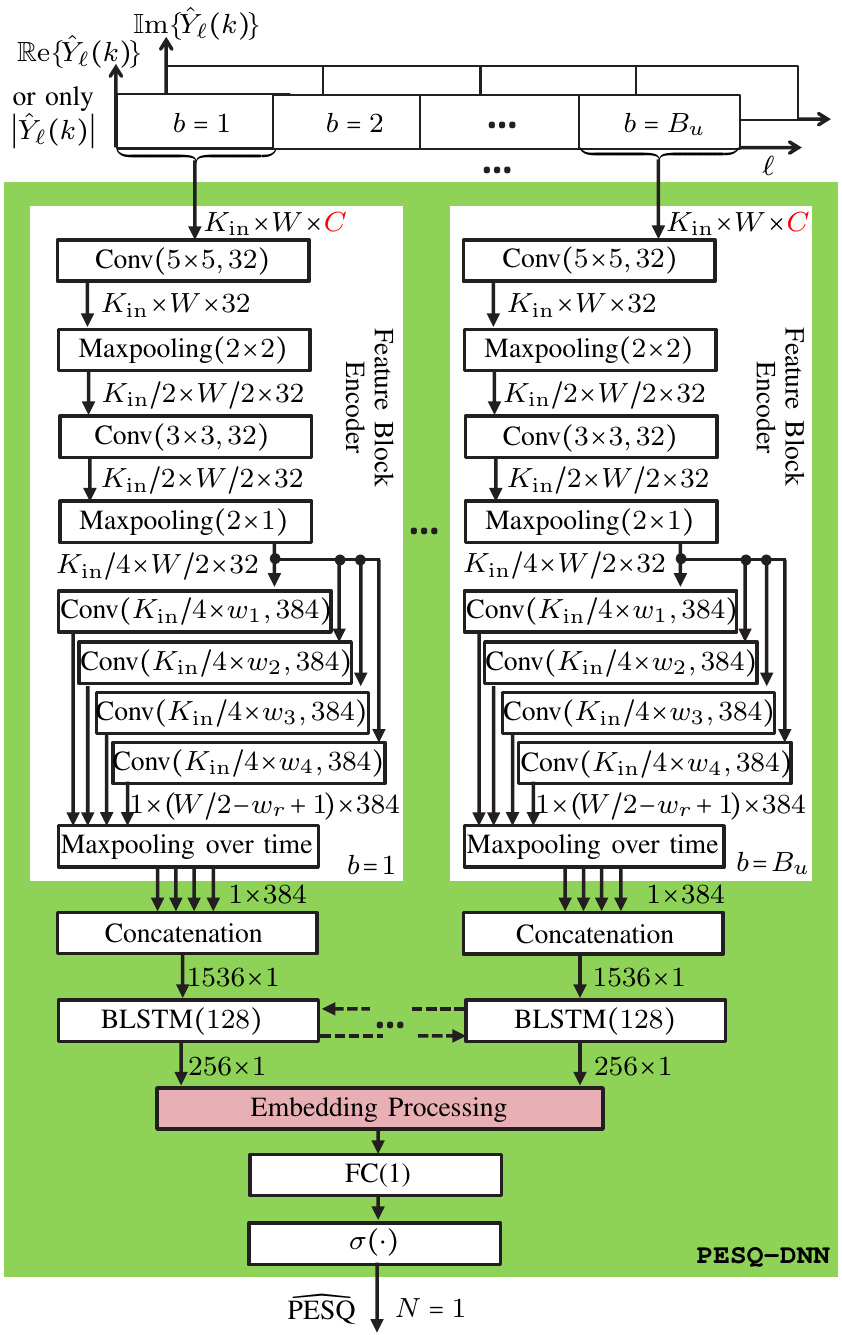}
	\caption{Topology of the proposed {\tt PESQ-DNN}. The number of input channels is set to either $C=1$ or $C=2$ for the {\tt PESQ-DNN} employing the amplitude or complex spectrogram as input, respectively. The ``Embedding Processing" model used in the baseline {\tt PESQNet} \cite{xu2021deepT} is recapitulated in Fig.\,\ref{old_embed}, while the employed model for our proposed {\tt PESQ-DNN} containing frame-level or block-level embeddings is illustrated in Fig.\,\ref{new_embed}.}
	\label{PESQNet}
\end{figure}
In this work, we propose a single end-to-end {\it non-intrusive} {\tt PESQ-DNN}, modeling ITU-T P862.2 PESQ but without \mbox{reference} input, to estimate PESQ scores of coded WB speech utterances in (wireless) speech communication systems. 
\subsection{{\tt PESQ-DNN} Topology}
Our proposed end-to-end {\it non-intrusive} {\tt PESQ-DNN} is shown in Fig.\,\ref{PESQNet}. The dimensions of the input and the output feature maps for each layer are depicted as {\it number of features} $\times$ {\it number of time frames} $\times$ {\it number of feature maps (if applicable)}. Firstly, we investigate two types of {\tt PESQ-DNN} employing either the amplitude or the complex spectrogram as input. For the first type of {\tt PESQ-DNN}, the input of the network is the amplitude spectrogram of the coded speech $\left|S_\ell(k)\right|$, with $\ell\!\in\!\mathcal{L}_u\!=\!\left \{1,2,\ldots,L_u\right \}$, and $L_u$ being the number of frames in the utterance with index $u$. Correspondingly, the number of input channels is set to $C=1$ in Fig.\,\ref{PESQNet}. On the other hand, the motivation for complex-valued input is straightforward: An algorithm employing the amplitude spectrogram as input cannot measure speech quality degradation caused by phase distortions. However, the original ITU-T P862.2 PESQ function considers phase influences in PESQ estimation to some extent, e.g., by estimating temporal delay between the samples of the degraded speech signal and its corresponding reference. The complex-valued input {\tt PESQ-DNN} employs the real and imaginary parts of the coded speech spectrogram, denoted as $\operatorname{\mathbb{R}e}\{\!\hat{Y}_\ell(k)\!\}$ and $\operatorname{\mathbb{I}m}\{\!\hat{Y}_\ell(k)\!\}$, respectively, as two separate input channels, resulting in $C=2$ in Fig.\,\ref{PESQNet}. For both types of {\tt PESQ-DNN}, the input is grouped into several feature blocks (feature matrices) indexed with $b\in\mathcal{B}_u=\left\{1, 2 \ldots, B_u\right\}$, with $B_u$ being the total number of blocks for utterance $u$. Each feature block has the same dimension $K_{\rm in}\!\times\! W\! \times\! C$, with $K_{\rm in}$ and $W$ being the number of input frequency bins and time frames per block, respectively.

Afterwards, the feature blocks are processed in parallel by identical subnetworks: A CNN-based encoder is employed to extract quality-related features from the input feature blocks. The 2D convolutional layers are denoted by Conv$(h\times w, f)$, with $f$ and $(h\times w\times \text{{\it number of input feature maps}})$ representing the number of filter kernels and the corresponding kernel size, respectively. In this CNN encoder, we use maxpooling layers with two different kernel sizes represented by $(2\times1\times \text{{\it number of input feature maps}})$ and $(2\times2\times \text{{\it number of input feature maps}})$. The extracted features are processed by a multi-width convolutional structure employing kernel widths of $w_1$, $w_2$, $w_3$, and $w_4$ to extract features with different time resolutions. The wide filters can capture long-term information, while the narrow ones focus on short-term information. The max-pooling-over-time layer chooses the highest activity along the time axis for each feature map. The subsequent concatenation delivers a feature vector with a fixed dimension to the BLSTM layer, which contains 128 neurons and is used to model temporal dependencies. Afterwards, we investigate various ``Embedding Processing" models (Fig.\,\ref{PESQNet}) to convert the per-utterance varying number of output embeddings of the BLSTM layer into a fixed length representation.

\begin{figure}[t!]
	\centering
	\includegraphics[width=0.49\textwidth]{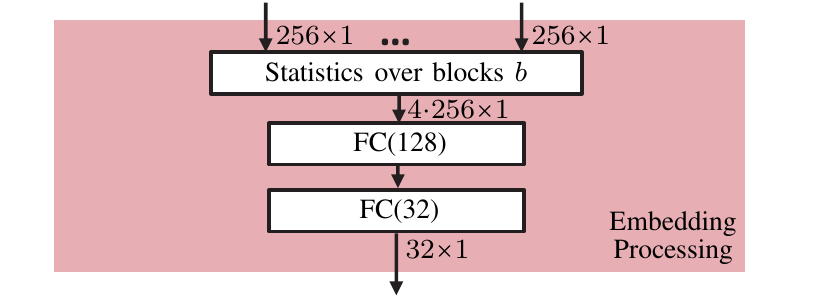}
	\caption{The employed ``Embedding Processing" model in Fig.\,\ref{PESQNet} for the {\tt PESQNet} proposed in \cite{xu2021deepT}. Four statistics (average, standard deviation, minimum, and maximum) over blocks $b$ are applied to the BLSTM outputs to deliver a feature vector with a fixed length.}
	\label{old_embed}
\end{figure}

The first investigated ``Embedding Processing" model from the original {\tt PESQNet} is shown in Fig.\,\ref{old_embed}, where four statistics (average, standard deviation, minimum, and maximum) over blocks $b$ are applied to the BLSTM outputs. Afterwards, the obtained feature vector has a fixed dimension of $4\cdot256 \times 1$ and is processed by two fully connected (FC) layers denoted as FC$(N)$, with $N$ being the number of output nodes. However, the baseline {\tt PESQNet} offers poor performance in PESQ estimation for coded speech utterances due to serious training instability issues. One can imagine that, e.g., packet losses will sparsely occur in some of the input blocks, which may dramatically change the value ranges of the BLSTM output belonging to these blocks. In consequence, there is a high dynamic range of the BLSTM outputs, even within the same utterance. Accordingly, the four statistics calculated from these BLSTM outputs are also highly unstable regarding the value range, finally resulting in training stability problems.

\begin{figure}[t!]
	\centering
	\includegraphics[width=0.49\textwidth]{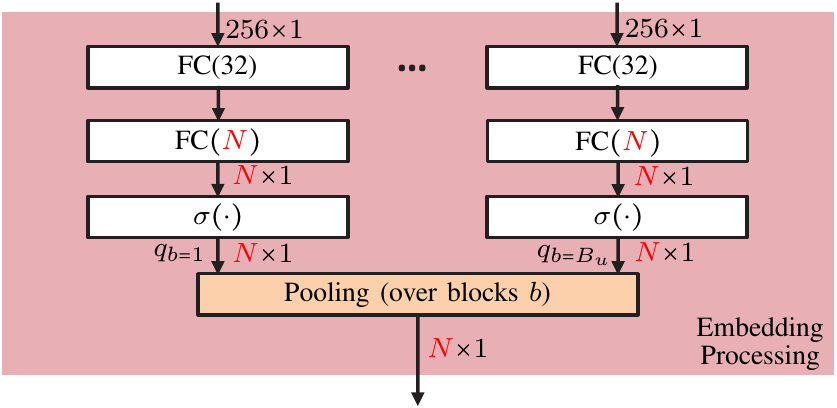}
	\caption{The employed ``Embedding Processing" model in Fig.\,\ref{PESQNet} for our proposed {\tt PESQ-DNN}. The ``Pooling" function is realized by either an average pooling or the attention pooling as illustrated in Fig.\,\ref{Attention}.}
	\label{new_embed}
\end{figure}

\begin{figure}[t!]
	\centering
	\includegraphics[width=0.49\textwidth]{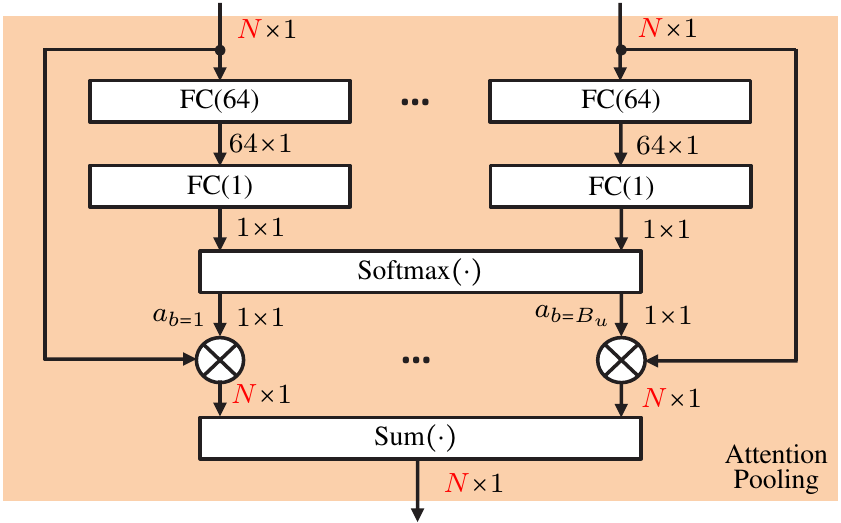}
	\caption{The employed ``Pooling" shown in Fig.\,\ref{new_embed}, here: attention pooling. We set $N=W$ and $N=1$ for the frame-level embeddings (FLE) and the block-level embeddings (BLE), respectively.}
	\label{Attention}
\end{figure}

Another cause of the training stability problem could be that without direct control or guidance, the BLSTM outputs have no physical meaning, thus, leading to a high dynamic value range. To mitigate this issue, we employ a modified ``Embedding Processing" model as shown in Fig.\,\ref{new_embed}, which is inspired by {\tt QualityNet} \cite{fu2018Inter}. The idea is to assign physical meanings to the intermediate embeddings, e.g., frame-level embeddings (FLE) PESQ scores or block-level embeddings (BLE) PESQ scores, even though the training target PESQ scores are still prepared utterance-wise. This intermediate FLE or BLE PESQ score is represented by $q_b$, with $b\in\mathcal{B}_u\!=\!\{1,2,...,B_u\}$. To estimate the FLE PESQ scores, we set $N=W$, representing that the number $N$ of neurons in the second FC layer equals the number $W$ of time frames in each block. Accordingly, we estimate the PESQ scores for each frame belonging to the corresponding input block. For the BLE PESQ scores, we set $N=1$ to output a single PESQ value for each block. The outputs of the FC layer are then processed by a gate function
\begin{equation} \label{gate}
\sigma(x)=3.6\cdot\text{sigmoid}(x)+1.04
\end{equation} 
to limit the range of the estimated FLE or BLE PESQ scores between $1.04$ and $4.64$, as it is with PESQ, determined by \mbox{ITU-T} P.862.2 \cite{ITUT_pesq_wb_corri}. Accordingly, if packet losses or noise occurs in one input block, its corresponding FLE or BLE PESQ scores will decrease within a limited range. A pooling layer is then employed to map the blocks' FLE or BLE PESQ scores to a single value, as shown in Fig.\,\ref{new_embed}. For the pooling layer, we investigate average and attention poolings, dubbed as ``AV" and ``AT", respectively, where the employed attention pooling is shown in Fig.\,\ref{Attention}. 

The idea of attention pooling is inspired by \cite{Vaswani2017a,mittag2021nisqa}, to explicitly consider the contributions of the FLE or BLE PESQ scores obtained from input blocks with poor speech quality or even speech pauses in the final PESQ estimation. As shown in Fig.\,\ref{Attention}, the attention score denoted as $a_b$, with $a_b \in \left[0,1\right]$ and $b\in\mathcal{B}_u$, is estimated for each block by the FC layers employing softmax as the output activation function. Subsequently, each block's FLE or BLE PESQ scores are multiplied by the estimated attention score and are summed over blocks, as shown in Fig.\,\ref{Attention}.

As shown in Fig.\,\ref{PESQNet}, the outputs obtained from either the embedding processing model in Fig.\,\ref{old_embed} or the proposed one in Fig.\,\ref{new_embed} are processed by the FC output layer with a single output node followed by the gate function \eqref{gate} to estimate the final utterance-related PESQ score.
\subsection{Training Loss Options}
The estimated PESQ score for the coded speech utterance should be as close as possible to its groundtruth PESQ score measured by ITU-T P.862.2 \cite{ITUT_pesq_wb_corri}. Accordingly, one option for training our proposed {\tt PESQ-DNN} is to employ the so-called {\it PESQ loss} defined as:
\begin{equation} \label{PESQ_Loss}
J^\text{PESQ}_u\!=\left(\widehat{\text{PESQ}}_u-\text{PESQ}_u\right)^2,
\end{equation}
with $\widehat{\text{PESQ}}_u$ and $\text{PESQ}_u$ being the estimated and the corresponding groundtruth PESQ scores measured by ITU-T P.862.2 \cite{ITUT_pesq_wb_corri}, respectively, and $u$ denoting the utterance index.

For training the {\tt PESQ-DNN} employing intermediate FLE PESQ scores, we introduce an additional constraint into the loss function \eqref{PESQ_Loss} as:
\begin{equation} \label{PESQ_FLE}
J^\text{PESQ}_u\!=\left(\widehat{\text{PESQ}}_u\!-\!\text{PESQ}_u\right)^2\!+\tfrac{\alpha_u}{B_u\cdot L_b}\!\sum\limits_{b\in\mathcal{B}_u}\sum\limits_{\ell\in\mathcal{L}_b}\!\left(q_{b}(\ell)\!-\!\text{PESQ}_u\right)^2,
\end{equation}
with $q_{b}(\ell)$ being the predicted intermediate {\it frame-level} PESQ scores for the block indexed with $b$, and frame index $\ell\!\in\!\mathcal{L}_b\!=\!\left \{1,2,\ldots,W\right \}$. Parameters $B_u=\left|\mathcal{B}_u\right|$ and $L_b\!=\!\left|\mathcal{L}_b\right|\!=\!W$ represent the total number of blocks for an utterance indexed with $u$ and the number of frames in each block, respectively. The utterance-wise weighting factor is represented by 
\begin{equation} \label{weights}
\alpha_u= 0.9^{\left|\text{PESQ}_u-\text{PESQ}_\text{max}\right|},
\end{equation}
with $\text{PESQ}_\text{max}=4.64$ being the maximum PESQ score defined in ITU-T P.862.2 \cite{ITUT_pesq_wb_corri}. The idea is that in loss function \eqref{PESQ_FLE}, the constraint reflected by the value of $\alpha_u$ should be higher for speech utterances with better quality: The intermediate FLE or BLE PESQ scores are then all encouraged to be equal to the utterance-wise PESQ. One can imagine that a speech utterance with a perfect overall perceptual quality should be annotated with the same good quality everywhere, i.e., the same high PESQ score in each block. Please note when $\alpha_u=0$, loss function \eqref{PESQ_FLE} reduces to the utterance-wise PESQ loss \eqref{PESQ_Loss}.

For the {\tt PESQ-DNN} employing intermediate BLE PESQ scores, the loss function \eqref{PESQ_FLE} is simplified to:
\begin{equation} \label{PESQ_BLE}
J^\text{PESQ}_u\!=\left(\widehat{\text{PESQ}}_u\!-\!\text{PESQ}_u\right)^2\!+\!\frac{\alpha_u}{B_u}\sum\limits_{b\in\mathcal{B}_u}\!\left(q_b\!-\!\text{PESQ}_u\right)^2,
\end{equation}
with $q_{b}$ being the intermediate {\it block-level} PESQ score for block indexed with $b$, see Fig.\,\ref{new_embed}.

For both loss functions \eqref{PESQ_FLE} and \eqref{PESQ_BLE}, the utterance-wise weighting factor $\alpha_u$ is calculated by \eqref{weights}.
%%%%%%%%%%%%%%%%%%%%%%%%%%%%%%%%%%%%%%%%%%%%%%%%%%%%%%%%%%%%%%%%%%%%%%%%%%%%%%%%%%%%%%%
\section{Experimental Setup and Databases}
\subsection{Databases and Preprocessing}
\begin{figure}[t!]
	\centering
	\includegraphics[width=0.49\textwidth]{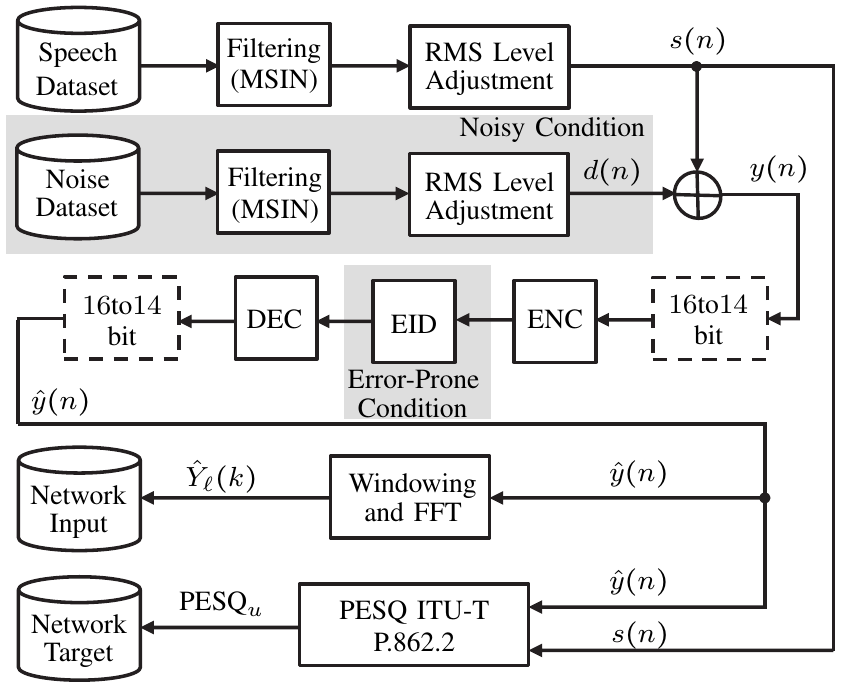}
	\caption{Training, validation, and test data processing for the proposed {\it PESQ-DNN} employing various codecs under clean, noisy, error-prone transmission, or tandeming conditions. The functions ``ENC" and ``DEC" represent the investigated codecs' encoder and decoder, respectively.}
	\label{Data_process}
\end{figure}
In this work, the speech signals used for training, development, and test have a sampling rate of $16\, \text{kHz}$ and are obtained from the NTT wideband speech database \cite{NTTML}, which contains $21$ languages represented by four female and four male speakers per language. Each speaker offers 12 speech utterances of 8s duration each. All the speech utterances in American English and German are used in our test set. Our training set is constructed with all speech utterances from the first three female and male speakers in all 19 other languages. Meanwhile, the development set uses all speech utterances from the remaining female and male speakers marked with ``{\tt f4}" and ``{\tt m4}" in the same 19 languages. Accordingly, the test performed in this work is completely speaker-independent and even partly language-independent: British English is one of $19$ languages used for training and development, while American English is used in the test.

The data preprocessing procedure is illustrated in Fig.\,\ref{Data_process}. In this work, the training, development, and test data is preprocessed based on the quality assessment plans for evaluating coded speech signals \cite{AMRWB_pro,ramo2005comparing,EVS7c,zhao2018convolutional}. The employed functions in Fig.\,\ref{Data_process} are taken from the ITU-T G.191 software tool library \cite{ITUT_G191}.

For our proposed {\tt PESQ-DNN}, we apply a periodic Hann window with a frame length of $512$ with a $50\%$ overlap in the ``Windowing and FFT" function shown in Fig.\,\ref{Data_process}. The obtained DFT spectrogram $\hat{Y}_\ell(k)$ is saved as the input for the {\tt PESQ-DNN}. The corresponding groundtruth PESQ targets for training are obtained from ITU-T P.862.2 \cite{ITUT_pesq_wb_corri}, employing the coded speech signal and its original reference speech signal as input, as shown in Fig.\,\ref{Data_process}.
\subsubsection{Training and Development Conditions}
The speech utterances are firstly processed by an MSIN high-pass filter \cite{ITUT_G191}. Afterwards, the active speech level of the filtered speech utterances measured by the root mean square (RMS) level is adjusted following ITU-T P.56 \cite{ITU56}. We target at three RMS levels for each speech utterance $s(n)$, namely $-36$, $-26$, and $-16$ dBov. To explicitly consider coded speech quality degradation under noisy conditions, the codec input signal $y(n)$ contains additive noise $d(n)$, see \eqref{mic}. The noise signals are taken from DEMAND \cite{thiemann2013diverse} and QUT \cite{dean2010qut}, comprising 35 different noise files shared in training and development. The additive noise signal is processed by the same MSIN high-pass filter used for the speech utterances, and the noise signal's RMS level is adjusted according to the desired signal-to-noise ratio (SNR). For each active speech level, we simulate $10\%$, $10\%$, and $80\%$ of the speech utterances with the SNR levels of $15$, $20$, and $\infty$ dB, respectively, where the SNR of $\infty$ dB represents the clean condition without additive noise. These relatively high SNR values take into account that in practice, the microphone-path signal processing employs a noise suppressor before speech encoding, typically achieving already a $10$...$20$ dB SNR improvement. 
 
After adjusting the active speech level of the speech utterances and mixing them with additive noises (if considered), the obtained transmitter-sided speech encoder input signal $y(n)$ is subject to coding. In Fig.\,\ref{Data_process}, the functions ``ENC" and ``DEC" represent the investigated codecs' encoder and decoder, respectively, comprising delay compensation for time alignment. In this work, we consider coded speech obtained from AMR-WB \cite{AMRWB} and EVS \cite{EVS} codecs operating at different bitrates in training and development. We employ the AMR-WB codec at seven bitrates (modes), including 6.6, 8.85, 14.25, 15.85, 18.25, 19.85, and 23.05 kbps, in the fixed-point implementation without DTX \cite{AMRWB_ANSI_C}. For the EVS codec, we employ its fixed-point implementation \cite{EVS_ANSI_C} at five bitrates, including 5.9, 8.0, 9.6, 16.4, and 24.4 kbps. Please note that EVS operating at 5.9 kbps requires DTX to be switched on, while the other modes are implemented without DTX.

The error insertion device (EID) in Fig.\,\ref{Data_process} is employed to insert frame losses into the bitstream obtained from ``ENC", simulating the error-prone transmission condition. In this work, we implement frame losses with EVS on clean speech utterances. We consider two types of frame losses, random and burst frame erasures. The random frame erasure is based on a Gilbert model \cite{mushkin1989capacity}, while the burst frame erasure employs the Bellcore model \cite{gilbert1960capacity}. Both types of frame erasures are measured by the frame error rate (FER), reflecting the ratio between the number of distorted frames and the number of all transmitted frames. For training and development, we employ FERs of $3\%$ and $10\%$, equally distributed among the two types of frame erasures, on $10\%$ of the {\it clean} speech utterances in each active speech level. Note that in case of a frame loss, the speech decoder (``DEC") provides means for error concealment and still outputs some estimated speech $\hat{y}(n)$.

In tandeming (TDM) transmission condition, the ``EID" is replaced by ``DEC" and the subsequent ``ENC," resulting in a serial concatenation of two WB codecs. Please note that in this work, additive noise and error-prone transmission are not considered in TDM evaluation, simply to keep the total number of experiments tractable. For training and development, speech utterances are firstly processed by G.722 \cite{ITUG722} and then by EVS \cite{EVS}, operating at 9.6 and 24.4 kbps. The function "16to14" in Fig.\,\ref{Data_process} is employed for bit conversion from 16 bits to 14 bits as required by AMR-WB and G.722 (in TDM condition). 

In this work, the training and development datasets are denoted as $\mathcal{D}^\text{train}$ and $\mathcal{D}^\text{dev}$, respectively. As explained above, the development set contains the same conditions as in training but with speech utterances from different speakers. We employ the development set $\mathcal{D}^\text{dev}$ not only for learning rate scheduling and early stopping during training. We also select the best {\tt PESQ-DNN} schemes based on the performance measured on the development set shown in Tab.\,\ref{dev_set}, which we will discuss later.
\subsubsection{Test Conditions}
The speech utterances used for the test are preprocessed in the same way as used for training, including MSIN filtering and RMS level adjustment. The active speech level of each test speech utterance is normalized to $-36$, $-26$, and $-16$ dBov. We explicitly consider the performance of our {\tt PESQ-DNN} on the coded speech obtained under clean, noisy, error-prone, and TDM transmission conditions.

In the {\it clean} condition, the signal path for the additive noise and the ``EID" function in Fig.\,\ref{Data_process} are deactivated and bypassed, respectively. We select $80\%$ of the test utterances in each {\it active speech level} for this clean test condition. The remaining $20\%$ of the test utterances will be used to prepare noisy speech utterances for the test in the {\it noisy} condition. For the clean test condition, we investigate coded speech utterances obtained from AMR-WB operating at 12.65 and 23.85 kbps and EVS operating at 7.2, 13.2, and 48 kbps. Please note that the employed operation modes for AMR-WB and EVS are unseen during training and development, enabling us to illustrate the codec mode generalization ability of the trained {\tt PESQ-DNN}. The performance of our {\tt PESQ-DNN} in the clean condition is shown in Tab.\,\ref{test_clean} and will be discussed later.

In the {\it noisy} condition, the signal path for the additive noise is activated while the ``EID" function in Fig.\,\ref{Data_process} is bypassed. We exclude the error-prone transmission condition and explicitly investigate the influence of additive noise on PESQ estimation for the coded speech. Please note that we report the noisy test condition for the EVS codec operating at 13.2 kbps. We take three {\it unseen} types of noise from the ETSI background noise database \cite{ETSI}, which differ from the used noise databases during training and development. The used unseen noise types include cafeteria noise, car noise with a speed of $100$ km/h, and traffic road noise. To prepare the speech utterances in noisy test condition, we use the remaining $20\%$ of the test utterances in each {\it active speech level} and mix them with the unseen types of noise signals to simulate an SNR of 15 dB. We will discuss the performance of our {\tt PESQ-DNN} in the noisy condition shown in Tab.\,\ref{test_noise} later.

In the {\it error-prone transmission condition}, the ``EID" function in Fig.\,\ref{Data_process} is switched on, while the noise signal path is deactivated, so we perform investigation in the clean condition. Furthermore, to keep the overall number of experiments tractable, we report the influence of the EID with the EVS codec operating at 13.2 and 48 kbps. To prepare the coded speech utterances impaired by the EID, we select $50\%$ of the test utterances in each {\it active speech level} and simulate an FER of $3\%$ and $6\%$ equally distributed among random and burst frame erasures. Please note that the FER of $6\%$ is {\it unseen} during training and development. 

In the {\it tandeming (TDM) transmission} condition, we employ two TDM cases on all test speech utterances: For the {\it seen} TDM case, the two codecs are concatenated in the same order as used in the training and development, where the speech utterances are firstly processed by G.722 and then by EVS at 13.2 kbps. For the {\it unseen} TDM case, we flip the order of the two codecs by firstly employing EVS at 13.2 kbps, followed by G.722. For both TDM transmission cases, we exclude the influences from additive noise and error-prone transmission.

In the following discussion, the employed test set will be denoted as $\mathcal{D}^\text{test}$. We report the performance of our {\tt PESQ-DNN} in each condition of $\mathcal{D}^\text{test}$ separately, as shown in Tabs.\,\ref{test_clean}, \ref{test_noise}, \ref{test_TDM}, \ref{test_EID}, and \ref{test_final}. We will give a detailed analysis of these tables in Section\,\uppercase\expandafter{\romannumeral 5}.
\subsection{Baselines}
Among prior art, two approaches \cite{fu2018Inter,catellier2020wawenets} directly predict PESQ, and another one can be easily adapted to predict PESQ \cite{reddy2021dnsmos}. These three methods will serve as baselines in our work {\it after retraining with our datasets}, thereby allowing direct performance comparison of the actual {\it network topologies} being used. The three approaches will be briefly sketched in the following.
\subsubsection{QualityNet DNN}
Fu et al.\!\cite{fu2018Inter} introduced the so-called {\tt QualityNet} to predict PESQ scores for speech signals degraded by additive noise or for enhanced speech signals obtained from a specific DNS model. The input of {\tt QualityNet} is the speech amplitude spectrogram. As one of the baselines in this work, we adopt the {\tt QualityNet} {\it DNN topology} to predict PESQ scores for coded speech by replacing the inputs and the corresponding targets during training. Accordingly, the topology of the {\tt QualityNet} baseline DNN is exactly the same as proposed in \cite{fu2018Inter}.

Similar to our proposed {\tt PESQ-DNN}, the {\tt QualityNet} baseline DNN employs a BLSTM layer as recurrent structure to deal with coded input speech signals of variable length. Please note that the loss functions \eqref{PESQ_FLE} and \eqref{PESQ_BLE}, which explicitly control the intermediate FLE and BLE PESQ scores during the {\tt PESQ-DNN} training, are inspired by the loss function used in {\tt QualityNet} training \cite{fu2018Inter}. Accordingly, by comparing to this baseline, we intend to make visible the benefits obtained merely from the network structure of the proposed {\tt PESQ-DNN}. This baseline is denoted as ``{\tt QualityNet} Baseline DNN \cite{fu2018Inter}" in the results tables.
\subsubsection{WaweNet DNN}
As another baseline, we adopt the {\it topology} of {\tt WaweNet} as proposed in \cite{catellier2020wawenets} to directly predict PESQ scores from the waveform of the coded speech signal. Similar to this work, the original {\tt WaweNet} proposed in \cite{catellier2020wawenets} predicts PESQ scores for coded speech signals from WB codecs but only considers the influence from additive background noise. Note that inverting the waveform of the speech signal by multiplying the original waveform with $-1$ will not influence perceptual quality or measured PESQ scores. This inverting invariance can be one of the training goals. For this purpose, we 
follow the authors of \cite{catellier2020wawenets} and
perform inverse phase augmentation (IPA) to all the training and development data used for {\tt WaweNet}. Furthermore, due to the lack of a recurrent layer, {\tt WaweNet} can only deal with speech signals of a fixed length of three seconds. In this work, a longer speech utterance is cut into several three-second segments without overlapping, and the remainder shorter than three seconds is discarded. The final estimated PESQ score for this long speech utterance equals the averaged estimations across all the segments. We call this baseline ``{\tt WaweNet} Baseline DNN \cite{catellier2020wawenets}" in this work.
\subsubsection{DNSMOS DNN}
Reddy et al.\ proposed a DNN-based measure called {\tt DNSMOS} \cite{reddy2021dnsmos} to predict human subjective rating scores following ITU-T P.808 \cite{ITU_P808} for enhanced speech signals obtained from DNS models. Its advanced version proposed in \cite{reddy2022dnsmos} offers separate estimations of human rating scores for the speech component, the background noise, and the overall enhanced speech qualities following ITU-T P.835 \cite{ITUT_P835}. Since we do not predict DNSMOS in this work, and also noise suppression algorithms are not in our focus, we adapt the {\tt DNSMOS} {\it DNN topology} proposed in \cite{reddy2021dnsmos} as a further baseline, but replace the input and the corresponding target during training. Please note that the adapted {\tt DNSMOS} DNN employs a log power spectrogram as input, with an FFT size of $K=320$. The {\tt DNSMOS} DNN proposed in \cite{reddy2021dnsmos} has a CNN-based structure and predicts human rating scores for input speech utterances with a fixed length of nine seconds. In this work, the used speech utterances having a length of eight seconds. Accordingly, for the training, development, and test of the adapted {\tt DNSMOS} baseline, all the speech utterances are zero-padded to nine seconds. This baseline is presented as ``{\tt DNSMOS} Baseline DNN \cite{reddy2021dnsmos}" in the following discussions.
\subsection{Performance Metrics}
Following \cite{abel2016instrumental,fu2018Inter} and \cite{reddy2021dnsmos}, the performance of the {\tt PESQ-DNN} and baseline models is measured by the mean absolute error (MAE)
\begin{equation} \label{MAE}
\text{MAE}=\frac{1}{U_\text{test}}\sum\limits_{u\in\mathcal{U}_\text{test}}\left|\widehat{\text{PESQ}}_u\!-\!\text{PESQ}_u\right|
\end{equation}
and the linear correlation coefficient (LCC) 
\begin{equation} \label{LCC}
\text{LCC}=\frac{\sum\limits_{u\in\mathcal{U}_\text{test}}\left(\widehat{\text{PESQ}}_u\!-\!\hat{\mu}\right)\!\cdot\!\left(\text{PESQ}_u\!-\!\mu\right)}{\sqrt{\sum\limits_{u\in\mathcal{U}_\text{test}}\left(\widehat{\text{PESQ}}_u\!-\!\hat{\mu}\right)^2}\!\cdot\!\sqrt{\sum\limits_{u\in\mathcal{U}_\text{test}}\left(\text{PESQ}_u\!-\!\mu\right)^2}}
\end{equation}
with $\widehat{\text{PESQ}}_u$ and $\text{PESQ}_u$ being the estimated and the corresponding groundtruth PESQ scores for speech utterance indexed with $u$, and $u\in\mathcal{U}_\text{test}$. In \eqref{MAE}, the total number of speech utterances in set $\mathcal{U}_\text{test}$ is denoted as $U_\text{test}=\left|\mathcal{U}_\text{test}\right|$. The mean values of the estimated and groundtruth PESQ scores over set $\mathcal{U}_\text{test}$ are represented by $\hat{\mu}$ and $\mu$, respectively.  Both MAE and LCC are calculated with the estimated PESQ score obtained from the {\tt PESQ-DNN} and its corresponding groundtruth measured according to ITU-T P.862.2 PESQ \cite{ITUT_pesq_wb_corri}. An accurately estimated PESQ score is reflected by a low MAE (ideally 0) and a high LCC (ideally 1.0).
\subsection{Training Setup}
For our proposed {\tt PESQ-DNN} and the {\tt QualityNet} baseline DNNs the inputs of the networks are normalized to zero-mean and unit-variance with statistics collected on the training dataset, as recommended by the proponents of {\tt QualityNet} in \cite{fu2019learning}. Meanwhile, for the {\tt WaweNet} baseline DNN and {\tt DNSMOS} baseline DNN, the corresponding waveform and log-power inputs are {\it not} normalized, thereby following \cite{catellier2020wawenets,reddy2021dnsmos}.

The number of input and output frequency bins in Fig.\,\ref{PESQNet} is set to $K_{\rm in}=260$. The last 3 frequency bins are redundant for compatibility with the two maxpooling operations in the employed {\tt PESQ-DNN} shown in Fig.\,\ref{PESQNet}. The widths of the convolutional kernels used in the {\tt PESQ-DNN} shown in Fig.\,\ref{PESQNet} are set to $w_i=2^{i-1}$, $i\in\left\{1,2,3,4\right\}$. The number of time frames in each feature block shown in Fig.\,\ref{PESQNet} is set to $W=16$. Except for the layers before the softmax (Fig.\,\ref{Attention}) or the gate function \eqref{gate} (Figs.\,\ref{PESQNet} and \ref{new_embed}), which employ a linear activation function, the rest of the layers in the {\tt PESQ-DNN} use the leaky rectified linear unit (ReLU) \cite{maas2013rectifier} as the activation function.

During {\tt PESQ-DNN} training, we employ the loss function \eqref{PESQ_Loss}, but alternatively \eqref{PESQ_FLE} and \eqref{PESQ_BLE} to explicitly control the intermediate FLE and BLE PESQ scores, respectively. The utterance-wise weighting factor $\alpha_u$ in loss functions \eqref{PESQ_FLE} and \eqref{PESQ_BLE} is calculated from \eqref{weights}. Please note that \eqref{PESQ_FLE} and \eqref{PESQ_BLE} are adapted from the loss used for the {\tt QualityNet} baseline training, in which the second loss term weighted by $\alpha_u$ is calculated by averaging over all frames belonging to the current input utterance without grouping them into blocks. The other baseline models are trained with the loss function \eqref{PESQ_Loss}, as it was proposed in their original works \cite{catellier2020wawenets,reddy2021dnsmos}.

For the {\tt PESQ-DNN} training, we employ a truncated backpropagation-through-time (BPTT) training scheme with a sequence length (unrolling depth of BPTT) equal to the total number of feature blocks $B_u$ belonging to the current input utterance with index $u$. A similar truncated BPTT scheme is used for training the {\tt QualityNet} baseline, but with a sequence length being the number of time frames belonging to the current input utterance. For all the trainings in this work, we employ the Adam optimizer with the initial learning rate of $10^{-4}$. The learning rate is multiplied by a factor of $0.6$ once the development loss measured on $\mathcal{D}^\text{dev}$ does not improve for two consecutive epochs. We stop the training after the development loss does not improve for six consecutive epochs, and the model providing the lowest development loss is saved. 
%%%%%%%%%%%%%%%%%%%%%%%%%%%%%%%%%%%%%%%%%%%%%%%%%%%%%%
\section{Experimental Results and Discussion}
\subsection{Ablation Studies for the New {\tt PESQ-DNN}}
\begin{table}[t!]
	\centering
	\caption{Ablation study on the proposed {\tt PESQ-DNN} performed on the {\bf development set} $\mathcal{D}^\text{dev}$. The {\tt PESQ-DNN} variants use either amplitude ($C=1$) or complex spectrograms ($C=2$) as input. The topologies of the {\tt PESQ-DNN} with block-level (BLE) or frame-level embeddings (FLE) are illustrated in Figs.\,\ref{PESQNet} and \ref{new_embed}. ``AV": average pooling, ``AT": attention pooling, see Fig.\,\ref{Attention}. Best results are in {\bf bold} font, and second best are \underline{underlined}. The best and the second-best {\tt PESQ-DNN} variants are selected and marked by ${\texg{\square}}$ and $\texg{\triangle}$, respectively.}
	\setlength\tabcolsep{2pt}
	\scriptsize
	\begin{tabular}{c c c c c c c c c c}
		\hline
		\multirow{2}{*}{\rotatebox{90}{Input\ }}&\multicolumn{1}{c}{\multirow{2}{*}{\stz{Method}}} & \multicolumn{4}{c}{\stz{MAE}} & \multicolumn{4}{c}{LCC} \\ \cmidrule(l{.25em}r{.25em}){3-6} \cmidrule(l{.25em}r{.25em}){7-10}
		\multicolumn{2}{l}{}                 & EVS & AWB & TDM & Total & EVS & AWB & TDM & Total \\ \hline
		\multirow{7}{*}{\rotatebox{90}{Amplitude\quad}} & \multicolumn{1}{l}{\stz{{\tt PESQNet} \cite{xu2021deepT}, $J \eqref{PESQ_Loss}$}} & 0.40 & 0.41 & 0.21 & 0.37 & 0.07 & 0.10 & 0.13 & 0.08 \\
		& \multicolumn{1}{l}{\stz{{\tt PESQ-DNN} (BLE), $J \eqref{PESQ_BLE}$, AV}} & 0.40 & 0.39 & 0.18 & 0.36 & 0.32 & 0.51 & 0.52 & 0.44 \\ 
		& \multicolumn{1}{l}{\stz{{\tt PESQ-DNN} (FLE), $J \eqref{PESQ_FLE}$, AV}} & 0.43 & 0.41 & 0.23 & 0.39 & 0.37 & 0.53 & 0.65 & 0.48 \\ 
		& \multicolumn{1}{l}{\stz{{\tt PESQ-DNN} (BLE), $J \eqref{PESQ_Loss}$, AV}} & 0.41 & 0.39 & 0.19 & 0.37 & 0.31 & 0.50 & 0.53 & 0.43 \\ 
		& \multicolumn{1}{l}{\stz{{\tt PESQ-DNN} (FLE), $J \eqref{PESQ_Loss}$, AV}} & 0.43 & 0.44 & 0.10 & 0.39 & 0.35 & 0.49 & 0.56 & 0.45 \\ 
		& \multicolumn{1}{l}{\stz{{\tt PESQ-DNN} (BLE), $J \eqref{PESQ_Loss}$, AT}} & 0.43 & 0.43 & 0.23 & 0.40 & 0.37 & 0.53 & 0.60 & 0.48 \\ 
		& \multicolumn{1}{l}{\stz{{\tt PESQ-DNN} (FLE), $J \eqref{PESQ_Loss}$, AT}} & 0.46 & 0.46 & 0.25 & 0.43 & 0.24 & 0.44 & 0.47 & 0.37 \\ \hline
		\multirow{7}{*}{\rotatebox{90}{Complex\quad}} & \multicolumn{1}{l}{\stz{{\tt PESQNet}, $J \eqref{PESQ_Loss}$}} & 0.41 & 0.41 & 0.21 & 0.38 & 0.08 & 0.11 & 0.12 & 0.08 \\
		& \multicolumn{1}{l}{\stz{{\tt PESQ-DNN} (BLE), $J \eqref{PESQ_BLE}$, AV}} & \underline{0.20} & \underline{0.12} & \underline{0.07} & \underline{0.14} &\underline{0.77} & {\bf 0.94} & {\bf 0.94} & {\bf 0.88} \\ 
		& \multicolumn{1}{l}{\stz{{\tt PESQ-DNN} (FLE), $J \eqref{PESQ_FLE}$, AV$^{\texg{\triangle}}$}} & {\bf 0.19} & \underline{0.12} & {\bf 0.06} & \underline{0.14} & {\bf 0.78} & {\bf 0.94} & {\bf 0.94} & {\bf 0.88} \\ 
		& \multicolumn{1}{l}{\stz{{\tt PESQ-DNN} (BLE), $J \eqref{PESQ_Loss}$, AV}} & {\bf 0.19} & \underline{0.12} & {\bf 0.06} & \underline{0.14} & {\bf 0.78} & \underline{0.93} & \underline{0.93} & \underline{0.87} \\ 
		& \multicolumn{1}{l}{\stz{{\tt PESQ-DNN} (FLE), $J \eqref{PESQ_Loss}$, AV}} & {\bf 0.19} & \underline{0.12} & {\bf 0.06} & \underline{0.14} & {\bf 0.78} & \underline{0.93} & \underline{0.93} & \underline{0.87} \\ 
		& \multicolumn{1}{l}{\stz{{\tt PESQ-DNN} (BLE), $J \eqref{PESQ_Loss}$, AT$^{\texg{\square}}$}} & {\bf 0.19} & {\bf 0.11} & {\bf 0.06} & {\bf 0.13} & {\bf 0.78} & {\bf 0.94} & \underline{0.93} & {\bf 0.88} \\ 
		& \multicolumn{1}{l}{\stz{{\tt PESQ-DNN} (FLE), $J \eqref{PESQ_Loss}$, AT}} & {\bf 0.19} & \underline{0.12} & {\bf 0.06} & \underline{0.14} & {\bf 0.78} & {\bf 0.94} & \underline{0.93} & {\bf 0.88} \\ \hline
	\end{tabular}
	\label{dev_set}
\end{table}
%%%%%%%%%%%%%%%%%%%%%%%%%%%%%%%%%%%%%%%%%%%%%%%%%%%%%%%%%
\begin{table*}[t!]
	\centering
	\caption{Performance in {\bf clean conditions}. Performance is measured on the {\bf test set} $\mathcal{D}^\text{test}$ and is reported separately on the coded speech obtained from EVS, AMR-WB (AWB) under different tested bitrates. Best results are in {\bf bold} font, and second best are \underline{underlined}.}
	\setlength\tabcolsep{3pt}
	\scriptsize
	\begin{tabular}{c c c c c c c c c c c c c c c c c c}
		\hline
		\multicolumn{2}{c}{\multirow{3}{*}{\stzh{Method}}} & \multicolumn{8}{c}{\stz{MAE}} & \multicolumn{8}{c}{\stz{LCC}}\\ \cmidrule(l{.25em}r{.25em}){3-10} \cmidrule(l{.25em}r{.25em}){11-18}
		\multicolumn{2}{l}{} & \multicolumn{3}{c}{\stz{AWB}} & \multicolumn{4}{c}{\stz{EVS}} & Total & \multicolumn{3}{c}{\stz{AWB}} & \multicolumn{4}{c}{\stz{EVS}} &Total\\ \cmidrule(l{.25em}r{.25em}){3-5} \cmidrule(l{.25em}r{.25em}){6-9} \cmidrule(l{.25em}r{.25em}){10-10}  \cmidrule(l{.25em}r{.25em}){11-13} \cmidrule(l{.25em}r{.25em}){14-17} \cmidrule(l{.25em}r{.25em}){18-18}
		\multicolumn{2}{r}{Bitrates (kbps):} & \stz{12.65} & \stz{23.85} & Avg. & \stz{7.2} &\stz{13.2}& \stz{48} & Avg. & - & \stz{12.65} & \stz{23.85} & Avg. & \stz{7.2} &\stz{13.2}& \stz{48} & Avg. & -  \\ \hline
		\multirow{3}{*}{\rotatebox{90}{REF$\quad$}} & \multicolumn{1}{l}{\stz{{\tt QualityNet} Baseline DNN} \cite{fu2018Inter}} & 0.16 & \underline{0.09} & \underline{0.13} & 0.30 & {\bf 0.08} & 0.29 & 0.25  & 0.19 & 0.79 & 0.83 & 0.74 & 0.52 & \underline{0.61} & {\bf 0.67} & 0.27  & 0.43 \\
		& \multicolumn{1}{l}{\stz{{\tt WaweNet} Baseline DNN} \cite{catellier2020wawenets}}& 0.14 & 0.15 & 0.15 & 0.14 & \underline{0.16} & 0.20 & \underline{0.17}  & \underline{0.16} & 0.76 & 0.69 & 0.77 & 0.51 & 0.52 & 0.27 & 0.84  & 0.82 \\
		& \multicolumn{1}{l}{\stz{{\tt DNSMOS} Baseline DNN} \cite{reddy2021dnsmos}}& 0.65 & 0.75 & 0.70  & 0.73 & 0.73 & 0.76 & 0.74  & 0.71 & \underline{0.80} & 0.75 & 0.77  & 0.48 & 0.59 & 0.21 & 0.83  & 0.81 \\ \hline
		\multirow{2}{*}{\rotatebox{90}{NEW}} & \multicolumn{1}{l}{\stz{{\tt PESQ-DNN} (FLE) $J \eqref{PESQ_FLE}$, AV$^{\texg{\triangle}}$}}& \underline{0.10} & {\bf 0.08} & {\bf 0.09} & {\bf 0.09} & {\bf 0.08} & \underline{0.10} & {\bf 0.09}  & {\bf 0.09}  & {\bf 0.84} & {\bf 0.89} & {\bf 0.89} & \underline{0.70} & {\bf 0.63} & \underline{0.61} & {\bf 0.94}  & {\bf 0.92} \\
		& \multicolumn{1}{l}{\stz{{\tt PESQ-DNN} (BLE) $J \eqref{PESQ_Loss}$, AT$^{\texg{\square}}$}} &  {\bf 0.09} & \underline{0.09} & {\bf 0.09} & \underline{0.10} & {\bf 0.08} & {\bf 0.09} & {\bf 0.09}  & {\bf 0.09}  & {\bf 0.84} & \underline{0.86} & \underline{0.86} & {\bf 0.73} & \underline{0.61} & 0.49 & \underline{0.93}  &  \underline{0.91} \\ \hline
	\end{tabular}
	\label{test_clean}
\end{table*}
In this section, we explicitly investigate which modifications introduced in our novel proposed {\tt PESQ-DNN} contribute most to stabilize the training process and to improve performance. Accordingly, reported in Tab.\,\ref{dev_set}, we perform a comprehensive ablation study on the trained variants of {\tt PESQ-DNN} by analyzing their performance measured on the development set $\mathcal{D}^\text{dev}$. Meanwhile, we will select the best two {\tt PESQ-DNN} variants as our proposed models based on their development set performance and use them for later comparison to the baselines on the test set $\mathcal{D}^\text{test}$. For each trained {\tt PESQ-DNN} variant, we report the averaged performance over the corresponding employed bitrates on the coded speech obtained from EVS, AMR-WB (AWB), and seen tandeming (TDM) condition (G.722 first, then EVS). As in the training set, EVS-coded speech from the development set considers influences from noisy and error-prone transmission conditions. {\tt PESQNet} \cite{xu2021deepT} and the complex-input {\tt PESQNet} employ loss function \eqref{PESQ_Loss}. {\tt PESQ-DNN} variants employing FLE and BLE shown in Fig.\,\ref{new_embed} are trained with either loss functions \eqref{PESQ_Loss}, \eqref{PESQ_FLE}, or \eqref{PESQ_BLE}. Models marked by ``AV" employ average pooling, while ``AT" represents attention pooling as illustrated in Fig.\,\ref{Attention}.

First, we investigate the influence of different input representations, amplitude or complex spectrograms, on the reference {\tt PESQNet} \cite{xu2021deepT} and on our proposed {\tt PESQ-DNN}, see also Fig.\,\ref{PESQNet} for $C=1$ or $C=2$. Concerning MAE \eqref{MAE}, in the upper half of Tab.\ \ref{dev_set} (amplitude spectra), there is no clear advantage visible of our new {\tt PESQ-DNN} variants vs.\,the reference {\tt PESQNet} \cite{xu2021deepT}. On LCC \eqref{LCC}, however, the novel {\tt PESQ-DNN} shows some improvement (in total: about $0.40$ vs.\ $0.08$). Still, LCC results of below $0.50$ are not good enough for a convincing solution. In the lower half of Tab.\ \ref{dev_set}, we see that the original {\tt PESQNet} adapted to complex-valued input shows actually the same poor performance as with the amplitude-input reference {\tt PESQNet} \cite{xu2021deepT} (total $\text{MAE}=0.38$, total $\text{LCC}=0.08$). Interestingly, however, our novel {\tt PESQ-DNN} takes a lot of profit from using complex spectrogram input, as can be easily seen by {\it significantly lower MAE results ($\approx 0.14$) and higher LCC results ($\approx 0.88$)}. Among the six {\tt PESQ-DNN} variants, on the other hand, we see no significant differences, neither w.r.t.\ BLE or FLE, nor w.r.t.\ the loss or pooling method. Accordingly, for the following experiments, we keep the two best methods (seven \nth{1} ranks and one \nth{2} rank, and six \nth{1} ranks and two \nth{2} ranks) marked by a green symbol, both employing complex spectral input. Note that the large step ahead in performance of our proposed {\tt PESQ-DNN} vs.\ the reference {\tt PESQNet} \cite{xu2021deepT} can be dedicated to the intermediate embeddings $q_b$ in Fig.\,\ref{new_embed} before pooling, which seems to be advantageous, independently whether the intermediate embeddings $q_b$ in Fig.\,\ref{new_embed} are explicitly controlled in training (losses \eqref{PESQ_FLE}, \eqref{PESQ_BLE}) or not (loss \eqref{PESQ_Loss}).

%%%%%%%%%%%%%%%%%%%%%%%%%%%%%%%%%%%%%%%%%%%%%%%%%%%%%%%%%%
\subsection{New {\tt PESQ-DNN} in Comparison to Baselines}
%%%%%%%%%%%%%%%%%%%%%%%%%%%%%%%%%%%%%%%%%%%%%%%%%%%%%%%%%%
\subsubsection{Clean Conditions}
First, we evaluate our proposed {\tt PESQ-DNN} variants and all baseline DNNs in the {\it clean condition} on $\mathcal{D}^\text{test}$, with coded speech signals obtained from EVS and AWB operating at different bitrates, as shown in Tab.\,\ref{test_clean}. To offer a better overview, we also report the {\it averaged} performance over all tested bitrates for AWB and EVS separately, as well as the {\it total} performance averaged {\it over all data} of tested conditions (not an average of MAE or LCC values!). Please note that influences from additive noise, error-prone transmission, and tandeming (TDM) are excluded in Tab.\,\ref{test_clean}. 

Among all methods, we observe that evaluated on MAE, the {\tt DNSMOS} baseline DNN is least suitable for predicting PESQ scores (total $\text{MAE}=0.71$), while its LCC is almost best among the baselines. On the other hand, among all methods, the {\tt QualityNet} baseline DNN has by far the poorest correlation (total $\text{LCC}=0.43$), thereby also not being suitable for PESQ prediction. In contrast, the {\tt WaweNet} baseline DNN shows quite good and balanced performance with a total MAE of $0.16$ and a total LCC of $0.82$, clearly on \nth{1} rank among the baselines. Please note, however, that our two new {\tt PESQ-DNN}s show an even better total performance than the baselines both in MAE and LCC metrics. They have equal or by far better performance than the baselines in each single condition, except one: for the 48 kbps EVS, our methods stay behind the (otherwise unbalanced) {\tt QualityNet} performance of $\text{LCC}=0.67$, but clearly excel at low EVS bitrate of 7.2 kbps ($\text{LCC}=0.70/0.73$ vs.\ $0.52$). In summary, reaching an $\text{MAE}=0.09$ vs.\ $0.16$, {\it our methods almost halve the MAE vs.\ the best of the baselines}, {\tt WaweNet}. {\it Concerning LCC, both of our methods reach values above $0.90$}, which we consider an encouraging result in comparison to the baselines. Note that since we adopted from {\tt QualityNet} \cite{fu2018Inter} the FLE-based intermediate embeddings $q_b$, accordingly, our strong correlation performance in this experiment cannot be deducted from the embeddings, but rather from the rest of our obviously advantageous {\tt PESQ-DNN} network structure. 
%\footnote{\texr{The PESQ algorithm is proven to have an average LCC of $0.94$ to the subjective MOS scores in ITU-T P.862.2 \cite{ITUT_pesq_wb_corri}, while POLQA superiors PESQ by offering an LCC of $0.97$ as specified in ITU-T P.863 \cite{ITUT_polqa_2018}. An instrumental quality measure for ABE tasks proposed by Abel et al.\ \cite{abel2016instrumental} shows an average LCC above $0.95$. (Need to discuss)}}.

\begin{figure}[t!]
	\centering
	\includegraphics[width=0.49\textwidth]{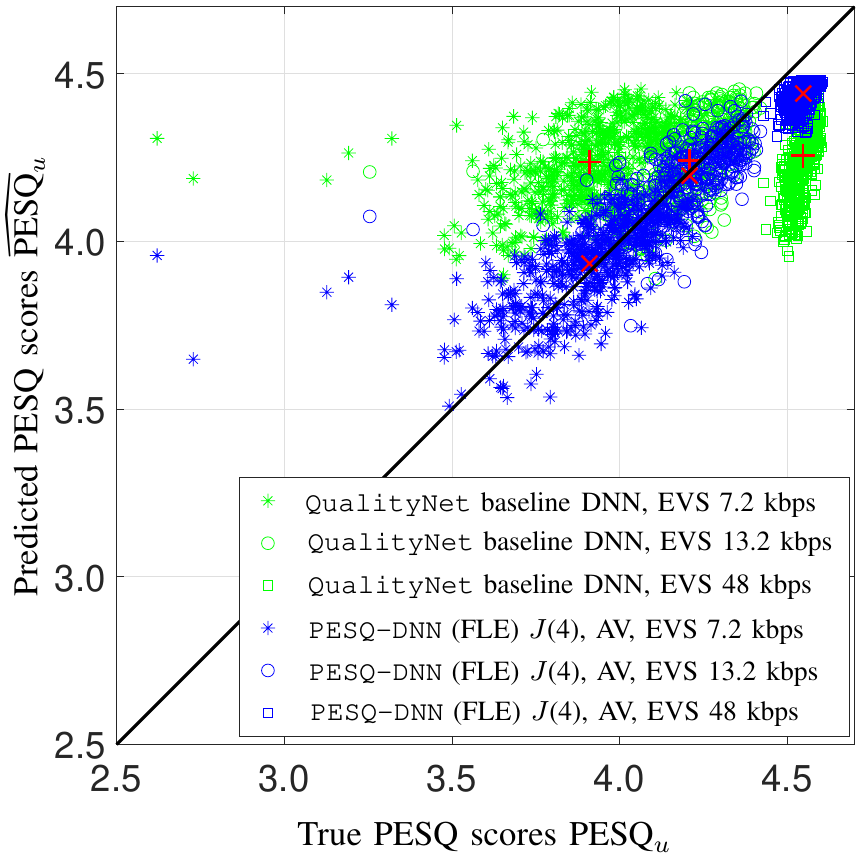}
	\caption{Scatter plot for the predicted PESQ scores $\widehat{\text{PESQ}}_u$ obtained from the strongest {\tt PESQ-DNN} variant and from the {\tt QualityNet} baseline DNN trained to predict PESQ, measured on {\bf EVS-coded} speech (7.2, 13.2, and 48 kbps) from $\mathcal{D}^\text{test}$ in {\bf clean conditions}. The markers $\texr{+}$ and $\texr{\times}$ represent the mean values of $\hat{\mu}^\text{\tt QualityNet}_j$ and $\hat{\mu}^\text{\tt PESQ-DNN}_j$ for $\widehat{\text{PESQ}}_u$ for a single-condition $\text{LCC}_j$ \eqref{LCC} with EVS mode $j\in\left\{7.2, 13.2, 48\right\}$ kbps, respectively (from left to right).}
	\label{clean_scatter}
\end{figure}

In Tab.\,\ref{test_clean}, we observe that individual codec mode LCC metrics of {\tt PESQ-DNN} are below $0.90$, but the average is above $0.90$. In contrary, for {\tt QualityNet} all codec-mode-individual LCC values are above the respective average. To analyze this observation, in Fig.\,\ref{clean_scatter} we present a scatter plot with the predicted PESQ scores $\widehat{\text{PESQ}}_u$ obtained from the strongest {\tt PESQ-DNN} variant employing FLE (blue) and the {\tt QualityNet} baseline (green) for EVS-coded speech, with its ground truth $\text{PESQ}_u$ measured by ITU-T P.862.2 \cite{ITUT_pesq_wb_corri}. Based on \eqref{LCC}, wee see that if changes of the ground truth $\text{PESQ}_u$ after subtracting its mean value $\mu$ are reflected by similar value changes of the predicted PESQ scores $\widehat{\text{PESQ}}_u$ after subtracting $\hat{\mu}$, we will obtain a high LCC score. Accordingly, this is reflected by the markers distributed along the diagonal line. To offer a better overview of the distributions, we plotted the mean values of $\hat{\mu}^\text{\tt QualityNet}_j$ and $\hat{\mu}^\text{\tt PESQ-DNN}_j$ at each individual EVS mode $j\in\left\{7.2, 13.2, 48\right\}$ kbps using the markers $\texr{+}$ and $\texr{\times}$, respectively, as shown in Fig.\,\ref{clean_scatter}. For both {\tt PESQ-DNN} and {\tt QualityNet}, the distribution of the markers belonging to individual EVS mode has a similar tendency to the diagonal line, reflected by all codec-mode-individual LCC values above $0.6$ in Tab.\,\ref{test_clean}. Meanwhile, the overall distribution could be reflected by the virtual line connected through the corresponding mean values of each EVS mode. Surprisingly, the codec-mode-individual mean values ($\texr{+}$) for {\tt QualityNet} are almost horizontally distributed, while the ones belonging to the proposed {\tt PESQ-DNN} variant ($\texr{\times}$) are scattered almost along the diagonal line. This explains why the average LCC for {\tt QualityNet} baseline measured on EVS-coded speech is much lower than its codec-mode-individual LCC value shown in Tab.\,\ref{test_clean}. The blue points are obviously distributed much closer to the diagonal, even though some outliers exist, while the green ones are scattered further away. This confirms that our proposed {\tt PESQ-DNN} offers a lower averaged MAE than the {\tt QualityNet} baseline in Tab.\,\ref{test_clean} (0.09 vs.\ 0.25).

\begin{table}[t!]
	\centering
	\caption{Performance in clean and {\bf unseen noisy conditions}. Performance is measured on EVS-coded speech at 13.2 kbps from {\bf test set} $\mathcal{D}^\text{test}$. Best results are in {\bf bold} font, and second best are \underline{underlined}.}
	\setlength\tabcolsep{2pt}
	\scriptsize
	\begin{tabular}{c c c c c c}
		\hline
		\multicolumn{2}{c}{\multirow{2}{*}{\stz{Method}}} & \multicolumn{2}{c}{\stz{Clean}} & \multicolumn{2}{c}{Noisy} \\ \cmidrule(l{.25em}r{.25em}){3-4} \cmidrule(l{.25em}r{.25em}){5-6}
		\multicolumn{2}{l}{}                 & \stz{MAE} & LCC & \stz{MAE} & LCC\\  \hline
		\multirow{3}{*}{\rotatebox{90}{REF$\quad$}} & \multicolumn{1}{l}{\stz{{\tt QualityNet} Baseline DNN} \cite{fu2018Inter}} & {\bf 0.08} & 0.61 & 1.54 & 0.23  \\ 
		& \multicolumn{1}{l}{\stz{{\tt WaweNet} Baseline DNN} \cite{catellier2020wawenets}} & \underline{0.16} & 0.52 & {\bf 0.26} & 0.57  \\
		& \multicolumn{1}{l}{\stz{{\tt DNSMOS} Baseline DNN} \cite{reddy2021dnsmos}} & 0.73 & 0.59 & \underline{0.28} & {\bf 0.78}  \\ \hline
		\multirow{2}{*}{\rotatebox{90}{NEW}} & \multicolumn{1}{l}{\stz{{\tt PESQ-DNN} (FLE), $J \eqref{PESQ_FLE}$, AV$^{\texg{\triangle}}$}} & {\bf 0.08} & {\bf 0.63} & 0.31 & \underline{0.75}  \\
		& \multicolumn{1}{l}{\stz{{\tt PESQ-DNN} (BLE), $J \eqref{PESQ_Loss}$, AT$^{\texg{\square}}$}} & {\bf 0.08} & \underline{0.62} & 0.29 & 0.72\\ \hline
	\end{tabular}
	\label{test_noise}
\end{table}

\subsubsection{Noisy Conditions}
We measure the performance of all the investigated methods in {\it unseen noisy conditions} on EVS-coded speech at 13.2 kbps from the test set $\mathcal{D}^\text{test}$. We report the performance separately on coded speech without and with additive noise, denoted as clean and noisy conditions in Tab.\,\ref{test_noise}, respectively. As we explore the influence of additive noise on PESQ estimation, we excluded TDM and error-prone transmission conditions in this experiment.

Among the baseline DNNs, the results give a mixed picture in the sense that the {\tt QualityNet} DNN is best in clean MAE and LCC, but worst in the noisy condition for both metrics. The {\tt DNSMOS} DNN, on the other hand, is best in noisy conditions, and w.r.t.\ MAE it performs even better in the noisy condition ($0.28$) than in the clean condition ($0.73$). However, compared to all other models, it performs by far poorest in MAE in the clean condition. Our proposed {\tt PESQ-DNN} variants, in contrast, show top performance in the clean condition, and in the noisy condition a balanced performance being only slightly behind top results. We note that our FLE variant with $J \eqref{PESQ_FLE}$ and average pooling (AV) turns out to be best.

%%%%%%%%%%%%%%%%%%%%%%%%%%%%%%%%%%%%%%%%%%%%%%%%%%%%%%%%%
\subsubsection{Tandeming Conditions}

In Tab.\,\ref{test_TDM}, we illustrate the performance of all investigated models in {\it TDM conditions}, where two WB codecs are serially concatenated. We exclude the influence from additive noise and error-prone transmission in this experiment. In the {\it seen} TDM condition, the order of the two concatenated codecs is the same as used in training and development, where G.722 at 64 kbps is followed by EVS at 13.2 kbps. In the {\it unseen} TDM condition, we flip the order of the two codecs by first employing EVS at 13.2 kbps, followed by G.722 operating at 64 kbps. 

The {\tt DNSMOS} and the {\tt QualityNet} baseline DNNs perform worse than both proposed {\tt PESQ-DNN} variants in any of the conditions and metrics. Interestingly, the {\tt WaweNet} baseline DNN shows decent quality both in seen and unseen TDM for both metrics, in the latter case it is even the strongest method w.r.t.\ MAE ($0.14$). Our proposed {\tt PESQ-DNN} variants, on the other hand, are by far strongest in seen TDM (both metrics) and also for LCC in unseen TDM condition. Note that the MAE in unseen TDM condition is strong (but not top-performing), and particularly note that both of our proposed methods again show a balanced performance. 

\begin{table}[t!]
	\centering
	\caption{Performance in {\bf seen} and {\bf unseen} {\bf tandeming (TDM) conditions}. Performance is measured on {\bf test set} $\mathcal{D}^\text{test}$. Best results are in {\bf bold} font, and second best are \underline{underlined}.}
	\setlength\tabcolsep{2pt}
	\scriptsize
	\begin{tabular}{c c c c c c c}
		\hline
		\multicolumn{2}{c}{\multirow{2}{*}{\stz{Method}}} & \multicolumn{2}{c}{\stz{Seen TDM}} & \multicolumn{2}{c}{Unseen TDM} \\ \cmidrule(l{.25em}r{.25em}){3-4} \cmidrule(l{.25em}r{.25em}){5-6}
		\multicolumn{2}{l}{}                 & \stz{MAE} & LCC & \stz{MAE} & LCC\\  \hline
		\multirow{3}{*}{\rotatebox{90}{REF$\quad$}} & \multicolumn{1}{l}{\stz{{\tt QualityNet} Baseline DNN} \cite{fu2018Inter}} & 0.18 & 0.48 & 0.28 & 0.41\\
		& \multicolumn{1}{l}{\stz{{\tt WaweNet} Baseline DNN} \cite{catellier2020wawenets}} & \underline{0.14} & 0.61 & {\bf 0.14} & 0.66 \\
		& \multicolumn{1}{l}{\stz{{\tt DNSMOS} Baseline DNN} \cite{reddy2021dnsmos}} & 0.72 & 0.67 & 0.63 & 0.76\\ \hline
		\multirow{2}{*}{\rotatebox{90}{NEW}} & \multicolumn{1}{l}{\stz{{\tt PESQ-DNN} (FLE) $J \eqref{PESQ_FLE}$ AV$^{\texg{\triangle}}$}} & {\bf 0.08} & \underline{0.79} & \underline{0.21} & {\bf 0.89}\\
		& \multicolumn{1}{l}{\stz{{\tt PESQ-DNN} (BLE) $J \eqref{PESQ_Loss}$, AT$^{\texg{\square}}$}} & {\bf 0.08} & {\bf 0.80} & 0.26 & \underline{0.85}\\ \hline
	\end{tabular}
	\label{test_TDM}
\end{table}
%%%%%%%%%%%%%%%%%%%%%%%%%%%%%%%%%%%%%%%%%%%%%%%%%%
\begin{table*}[t!]
	\centering
	\caption{Performance in {\bf error-prone conditions} with different frame error rates (FER). Performance is measured on EVS-coded speech at 13.2 and 48 kbps from {\bf test set} $\mathcal{D}^\text{test}$. Best results are in {\bf bold} font, and second best are \underline{underlined}.}
	\setlength\tabcolsep{3pt}
	\scriptsize
	\begin{tabular}{c c c c c c c c c c c c c c c c}
		\hline
		\multicolumn{2}{c}{\multirow{3}{*}{\stzh{Method}}} & \multicolumn{6}{c}{\stz{EVS @ 13.2 kbps}} & \multicolumn{6}{c}{\stz{EVS @ 48 kbps}} & \multicolumn{2}{c}{\stz{Total (FER 3\%/6\%)}}\\ \cmidrule(l{.25em}r{.25em}){3-8} \cmidrule(l{.25em}r{.25em}){9-14} \cmidrule(l{.25em}r{.25em}){15-16}
		\multicolumn{2}{l}{} & \multicolumn{3}{c}{\stz{MAE}} & \multicolumn{3}{c}{\stz{LCC}} & \multicolumn{3}{c}{\stz{MAE}} & \multicolumn{3}{c}{\stz{LCC}} & \multicolumn{1}{c}{\stz{MAE}} & \multicolumn{1}{c}{\stz{LCC}}\\ \cmidrule(l{.25em}r{.25em}){3-5} \cmidrule(l{.25em}r{.25em}){6-8} \cmidrule(l{.25em}r{.25em}){9-11} \cmidrule(l{.25em}r{.25em}){12-14} \cmidrule(l{.25em}r{.25em}){15-15} \cmidrule(l{.25em}r{.25em}){16-16}
		\multicolumn{2}{r}{FER:} & \stz{0\%} & 3\% & 6\% & \stz{0\%} & 3\% & 6\% & \stz{0\%} & 3\% & 6\% & \stz{0\%} & 3\% & 6\% & - & -\\ \hline
		\multirow{3}{*}{\rotatebox{90}{REF$\quad$}} & \multicolumn{1}{l}{\stz{{\tt QualityNet} Baseline DNN} \cite{fu2018Inter}} & {\bf 0.08} & {0.38} & 0.70 & 0.61 & {0.64} & 0.51 & 0.29 & {\bf 0.23} & \underline{0.42} & {\bf 0.67} & 0.53 & 0.60 & 0.43 & 0.54\\
		& \multicolumn{1}{l}{\stz{{\tt WaweNet} Baseline DNN} \cite{catellier2020wawenets}} & \underline{0.16} & {\bf 0.27} & \underline{0.49} & {0.52} & 0.54 & 0.46 & {0.20} & \underline{0.29} & 0.47 & {0.27} & 0.44 & 0.42 & \underline{0.38} & 0.67\\
		& \multicolumn{1}{l}{\stz{{\tt DNSMOS} Baseline DNN} \cite{reddy2021dnsmos}} & 0.73 & 0.39 & {\bf 0.33} & 0.59 & 0.57 & 0.42 & 0.76 & 0.35 & {\bf0.24} & 0.21 & 0.56 & 0.41 & {\bf 0.33} & 0.61 \\ \hline
		\multirow{2}{*}{\rotatebox{90}{NEW}} &\multicolumn{1}{l}{\stz{{\tt PESQ-DNN} (FLE) $J \eqref{PESQ_FLE}$, AV$^{\texg{\triangle}}$}} & {\bf 0.08} & \underline{0.32} & 0.54 & {\bf 0.63} & {\bf 0.67} & \underline{0.63} & \underline{0.11} & 0.32 & 0.51 & \underline{0.61} & {\bf 0.76} & {\bf 0.80} & 0.45 & {\bf 0.75}\\
		&\multicolumn{1}{l}{\stz{{\tt PESQ-DNN} (BLE) $J \eqref{PESQ_Loss}$, AT$^{\texg{\square}}$}} & {\bf 0.08} & 0.33 & 0.56 & \underline{0.62} & \underline{0.66} & {\bf 0.66} & {\bf 0.09} & 0.33 & 0.53 & {0.49} & \underline{0.68} & \underline{0.70} & 0.46 & \underline{0.74} \\\hline
	\end{tabular}
	\label{test_EID}
\end{table*}
\subsubsection{Error-Prone Transmission Conditions}
We present the performance of all investigated models concerning frame loss in error-prone transmission conditions in Tab.\,\ref{test_EID}. To better investigate the influence of frame losses on PESQ estimation, we exclude the noisy and TDM transmission conditions in this experiment. We report the performance separately on EVS-coded speech at 13.2 and 48 kbps with FERs of $0\%$, $3\%$, and $6\%$, as shown in Tab.\,\ref{test_EID}. Among the employed FERs, an FER of $6\%$ is unseen during training and development, and an FER of $0\%$ represents error-free transmission conditions, thus offering the same performance as in clean conditions shown in Tab.\,\ref{test_clean}. Furthermore, each simulated FER contains equally distributed random and burst frame erasures. To offer a better overview of all investigated models in error-prone transmission conditions, we present the total performance averaged over all the employed bitrates only with FERs of $3\%$ and $6\%$, as shown in Tab.\,\ref{test_EID}.

The results show that the error-prone transmission condition was a tough case for all investigated models. For most of them, the MAE results under frame errors are much higher than for FER $=0\%$. Surprisingly, the {\tt DNSMOS} baseline DNN has lower MAE in error-prone condition as it has in the FER $=0\%$ case. This is a similar observation to Tab.\,\ref{test_noise}, where we observed a better noisy condition MAE as a clean condition MAE for the {\tt DNSMOS} network. We can conclude that it has topological strengths in harsh conditions (only). Note that while our proposed methods fall a bit behind the baselines w.r.t.\ MAE, they are well ahead when it comes to correlation (LCC): Overall, our proposed methods are \nth{1} and \nth{2} ranked as concerns LCC, while having decent MAE performance. 

\subsubsection{Overall Performance}
\begin{table}[t!]
	\centering
	\caption{{\bf Overall performance} averaged over all tested conditions ("total") and over all conditions but without the error-prone transmission condition ("total without EID"). Performance is measured on {\bf test set} $\mathcal{D}^\text{test}$. Best results are in {\bf bold} font, and second best are \underline{underlined}.}
	\setlength\tabcolsep{2pt}
	\scriptsize
	\begin{tabular}{c c c c c c}
		\hline
		\multicolumn{2}{c}{\multirow{2}{*}{\stz{Method}}} & \multicolumn{2}{c}{\stz{Total}} & \multicolumn{2}{c}{\stzh{\makecell{\stz Total \\ without EID}}} \\ \cmidrule(l{.25em}r{.25em}){3-4} \cmidrule(l{.25em}r{.25em}){5-6}
		\multicolumn{2}{l}{}&                 \stz{MAE} & LCC & $\text{MAE}^*$ & $\text{LCC}^*$ \\ \hline
		\multirow{3}{*}{\rotatebox{90}{REF$\quad$}} & \multicolumn{1}{l}{\stz{{\tt QualityNet} Baseline DNN} \cite{fu2018Inter}} &  0.27 & 0.67 & 0.24 & 0.44 \\ 
		& \multicolumn{1}{l}{\stz{{\tt WaweNet} Baseline DNN} \cite{catellier2020wawenets}} & \underline{0.19} & 0.79 & 0.16 & 0.87 \\
		& \multicolumn{1}{l}{\stz{{\tt DNSMOS} Baseline DNN} \cite{reddy2021dnsmos}} & 0.63  & 0.78 & 0.67 & 0.88 \\ \hline
		\multirow{2}{*}{\rotatebox{90}{NEW}} &\multicolumn{1}{l}{\stz{{\tt PESQ-DNN} (FLE) $J \eqref{PESQ_FLE}$, AV$^{\texg{\triangle}}$}} & {\bf 0.16} & {\bf 0.84} & {\bf 0.11} & {\bf 0.92} \\
		& \multicolumn{1}{l}{\stz{{\tt PESQ-DNN} (BLE) $J \eqref{PESQ_Loss}$, AT$^{\texg{\square}}$}} & {\bf 0.16} & \underline{0.82} & \underline{0.12} & \underline{0.90} \\ \hline
	\end{tabular}

	\label{test_final}
\end{table}
Finally, we report the overall test performance of the investigated methods in Tab.\,\ref{test_final}. The measurements are calculated on the coded speech signals obtained from both EVS and AWB codecs operating at different bitrates and considering all the tested conditions illustrated in Tabs.\,\ref{test_clean}, \ref{test_noise}, \ref{test_TDM}, and \ref{test_EID}. Since most of the investigated models show a bit limited performance in the error-prone condition in Tab.\,\ref{test_EID}, we also report the overall performance excluding the error-prone condition.

In total, taking the data (not the MAE or LCC) over all conditions, both of our proposed methods take on the \nth{1} and the \nth{2} ranks, respectively. Leaving out the data for the error-prone transmission condition, again both of our methods are \nth{1} and \nth{2} ranked. In this case, the top-performing method "{\tt PESQ-DNN} (FLE) $J \eqref{PESQ_FLE}$, AV$^{\texg{\triangle}}$" achieves a strong correlation of $0.92$ and an MAE of only $0.11$, thereby marking our final proposed method for non-intrusive PESQ prediction by a DNN.

%%%%%%%%%%%%%%%%%%%%%%%%%%%%%%%%%%%%%%%%%%%%%%%%%%%%%%%%%%%%
\section{Conclusions}
This work proposed an end-to-end non-intrusive {\tt PESQ-DNN} to evaluate coded speech quality in speech communication systems. We illustrated the potential of our proposed {\tt PESQ-DNN} by non-intrusively estimating the perceptual evaluation of speech quality (PESQ) scores of the wideband coded speech obtained from AMR-WB or EVS codecs operating at different bitrates in noisy, tandeming, and error-prone transmission conditions. We compare our methods with the state-of-the-art networks of {\tt QualityNet}, {\tt WaweNet}, and the {\tt DNSMOS} DNN applied to PESQ estimation by measuring the mean absolute error (MAE) and the linear correlation coefficients (LCC). As a core contribution of this work, we performed a comprehensive ablation study illustrating the influence and value of the introduced modifications to the proposed {\tt PESQ-DNN}, including the input types of the neural network, variations of the {\tt PESQ-DNN} topology in terms of different embedding processing models and loss functions employed during training. Averaged over all tested conditions, the proposed {\tt PESQ-DNN} offers the best total MAE and LCC of 0.11 and 0.92, respectively, without considering frame losses. Also under frame loss, its LCC is ahead of any of the baselines. Note that our network could be similarly used to non-intrusively predict POLQA or other intrusive metrics.

\ifCLASSOPTIONcaptionsoff
%\newpage
\fi

%\clearpage
%
\bibliographystyle{IEEEtran}
\bibliography{mainTrans21}

%\end{IEEEbiography}
%% insert where needed to balance the two columns on the last page with
%%\newpage


%\vfill


\end{document}